\documentclass[twoside,final,9pt,letterpaper,twocolumn]{article}

\usepackage{amsmath}
\usepackage{amssymb}
\usepackage{amstext}
\usepackage{alltt}
\usepackage{amsthm}
\usepackage{alltt}
\usepackage{float}

\newif\ifpdf
\ifx\pdfoutput\undefined
\pdffalse 
\else
\pdfoutput=1 
\pdftrue
\fi
\ifpdf
\usepackage[pdftex]{graphicx}
\else
\usepackage{graphicx}
\fi

\newtheorem{theorem}{Theorem}
\newtheorem{definition}{Definition}

\newtheorem{example}{Example}

\newcommand{\ket}[1]{\ensuremath{| #1\rangle}}
\newcommand{\bra}[1]{\ensuremath{\langle #1|}}

\newcommand{\Z}{\mathbb{Z}}

\newcommand{\C}{\mathbb{C}}

\newcommand{\tr}{\mathrm{tr}}

\floatstyle{boxed}
\newfloat{boxfloat}{h}{lop}[section]
\floatname{boxfloat}{Inset}

\begin{document}

\ifpdf
\DeclareGraphicsExtensions{.pdf, .jpg, .tif}
\else
\DeclareGraphicsExtensions{.eps, .jpg}
\fi

\title{
   Models of  Quantum Cellular Automata}
\author{ Carlos A. P\'{e}rez-Delgado and Donny Cheung}
\date{Institute for Quantum Computing,\\
University of Waterloo,\\
Waterloo, ON N2L 3G1, Canada}    

\maketitle

\begin{abstract} In this paper we present a systematic view of Quantum Cellular Automata (QCA), a mathematical formalism of quantum computation. First we give a general mathematical framework with which to study QCA models. Then we present four different QCA models, and compare them. One model we discuss is the traditional QCA, similar to those introduced by Shumacher and Werner, Watrous, and Van Dam. We discuss also Margolus QCA, also discussed by Schumacher and Werner. We introduce two new models, Coloured QCA, and Continuous-Time QCA. We also compare our models with the established models. We give proofs of computational equivalence for several of these models. We show the strengths of each model, and provide examples of how our models can be useful to come up with algorithms, and implement them in real-world physical devices.
\end{abstract}

\section{Introduction}

Quantum cellular automata (QCA) research has seen significant growth in the
recent years. This model of computation has been appearing in the literature,
sometimes with different names, and in several different guises. Quantum
lattice gases, pulse-driven quantum computers, and translation-invariant
quantum operators, are all instances of QCA, and yet little has been discussed
of the relationship between these.

The purpose of this paper is to give one unifying view of all QCA type
phenomena, in the form of a model of computation. We intend to show how all
previous models, whether given the name of QCA or some other, are either
instances of, or equivalent to, our model of QCA. Along the way, we will
survey some of the major work in the field of QCA.

\section{Classical Cellular automata}

We start our discussion with a short review of classical cellular automata.

A \emph{Cellular Automaton} (CA) consists of a lattice structure, where each
cell is in one of a finite number of predetermined cell states.  At each
discrete time-step, every cell is updated, in parallel, according to a local,
spatially uniform rule.  This gives us a model of computation which is
different from the usual circuit model or the Turing machine model.

\begin{figure}
\begin{center}
    \includegraphics[scale=0.9]{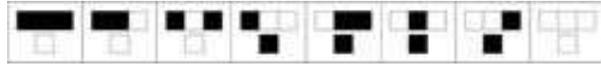}
\end{center}

\caption{\label{rule30} \emph{Wolfram's ``Rule 30'':} Each cell has a binary
state, and it is updated according to the value of itself, and its two
nearest neighbours, one to each side. There are eight possible states for the 
\emph{neighbourhood} of any particular cell.  The diagram above gives the
possible neighbourhood configurations, with the corresponding updated state
for the centre cell below.  Reading the second row as binary digits, one
obtains Wolfram's name for this rule: 30. }
\end{figure}

\begin{figure}
\begin{center}
    \includegraphics[scale=1.0]{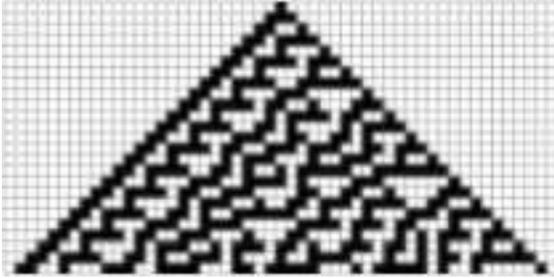}
\end{center}
\caption{\label{rule30ev} \emph{Rule 30 in motion:} Here we can see several
steps of the CA Rule 30 as presented in Figure \ref{rule30}. This is a
one-dimensional lattice, with the downward vertical dimension representing
successive time steps.}
\end{figure}

An example of a CA rule is given in figure \ref{rule30}. In brief, a
CA is a state lattice with an update rule that updates the value of each
lattice cell at each time step. The rule is local, in that the updated state
for an individual cell depends only on the current states of the cell itself
and of its neighbour cells.  The set of cells on which the updated value of
cell $x$ depends is called the \emph{neighbourhood} of $x$. The size and form
of the neighbourhood may vary from CA to CA.

We can see the time evolution of the rule in Figure \ref{rule30ev}.

\begin{figure}
 \begin{center}
  \includegraphics[scale=0.4]{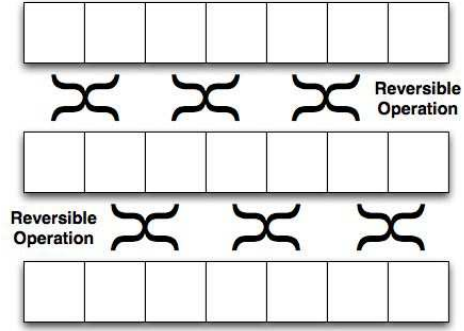}
  \end{center}

\caption{\label{margolus1} \emph{Margolus Partitioning CA Scheme}}
\end{figure}

In order to quantize the CA model, it is desirable to first make the model
reversible.  On the other hand, reversible computing predates quantum computing by several
decades.  One way to ensure that a CA is reversible is to use a
\emph{Margolus} partitioning scheme, in which the lattice is divided into
tiles. A reversible operation is performed on each tile. On consecutive
time-steps the tiling is staggered to allow the possibility of data
propagation (see Figure \ref{margolus1}).

\subsection{Quantum Cellular Automata}

We wish to to formulate a model for \emph{quantum} cellular automata.  Several
properties are desirable:

\begin{itemize}
\item The model should be a generalization of the classical CA and subsume the latter as a special case.
\item The model should allow for quantum computation; it should be equivalent to the quantum circuit model.
\item The model should ideally be reversible.
\item The model should be a natural abstraction for quantum computation on particular realistic devices.
\end{itemize}

With this in mind, we first present a generic, basic model which will be our starting point for all our QCA studies.

\subsubsection{The Basic Model}

In our quantum cellular automaton model, we consider qubits arranged in an
integer lattice of dimension $d$, $\mathbb{Z}^d$.

Instead of trying to define an state on an infinite lattice, we define a state
$\rho$ as a family of states on finite subsets of the lattice.  For each
finite $A\subseteq\mathbb{Z}^d$, we take $\rho_A$ to be a state in the Hilbert
space $\mathcal{H}^{\otimes A}$ over the subset $A$.  The family of states
$\rho=\{\rho_A\}_{A\subseteq\mathbb{Z}^d,A\textrm{ finite}}$ must also satisfy
a consistency condition.  For every finite subset $A\subseteq\mathbb{Z}^d$,
given $B\subseteq A$, we must have
$$\rho_B=\tr_{A\setminus B}(\rho_A).$$
In other words, partial traces must be consistent.

For a QCA, we must also have a global evolution operator.  However, for any
particular qubit in the lattice, interactions in one time-step must be
localized to within a given neighbourhood of the qubit.  To this end, we will
first define a \emph{neighbourhood scheme} as a finite set $N$ of elements of
$\mathbb{Z}^d$ which includes the zero vector $0$, and then for each given
lattice position $x$, we define the the neighbourhood of $x$ to be
$\mathcal{N}(x)=x+N$.  For a finite subset $A\subseteq\mathcal{Z}^d$, we will
define $\mathcal{N}(A)$ to be the union of the neighbourhoods of the elements
of $A$.  Naturally, neighbourhood sets are translation independent, that is,
given a lattice translation $\tau$, we have
$\mathcal{N}(\tau(x))=\tau(\mathcal{N}(x))$.

Now, given a finite subset $A\subseteq\mathbb{Z}^d$, the result of a
global evolution operator can be determined from its action on the subset
$\mathcal{N}(A)$.  Accordingly, we define our global evolution operator $U$ as
a family of unitary operators $U_{\mathcal{N}(A)}$ acting on neighbourhoods
for every finite subset $A\subseteq\mathbb{Z}^d$.  The action of $U$ on a
given state $\rho$ is given by
$$(U\rho)_A=\tr_{\mathcal{N}(A)\setminus A}\left(U_{\mathcal{N}(A)}
\rho_{\mathcal{N}(A)}U^{\dagger}_{\mathcal{N}(A)}\right).$$
Of course, we also require $U$ to satisfy translation independence, so that
for any subset $A\subseteq\mathbb{Z}^d$ and lattice translation $\tau$, we have
$U_{\mathcal{N}(\tau(A))}=\tau(U_{\mathcal{N}(A)})$.

\subsubsection{Deriving Global and Local Operators}

We consider the operator $U$, consisting of a unitary operator
$U_{\mathcal{N}(A)}$ for each
finite subset $A\subseteq\mathbb{Z}^d$, as the global evolution operator for a
given cellular automaton.  From this, it would be desirable to find a local
operator, which could reciprocally be used to derive the global operator.

The unitary operator $U_{\mathcal{N}(x)}$, corresponding to the subset $\{x\}$
consisting of a single element, makes a natural choice for a local operator.
However, deriving the global operator from this requires some work, as we
require a family of unitary operators which satisfy the consistency condition
for each finite subset of $\mathbb{Z}^d$.

To this end, for the qubit at lattice position $x$, we define $S_x^0$ to be
the set of pure states $\ket{\psi}$ over the lattice subset
$\mathcal{N}(x)$ which satisfy
$$\tr_{\mathcal{N}(x)\setminus\{x\}}
\left(U_{\mathcal{N}(x)}\ket{\psi}\!\bra{\psi}U_{\mathcal{N}(x)}^{\dagger}
\right)=\ket{0}\!\bra{0},$$
and similarly, we define $S_x^1$ to be the set of pure states which map to 
$\ket{1}\!\bra{1}$.  In general, the set $S_x^j$ is an affine space for which
a basis may be computed for any given $U_{\mathcal{N}(x)}$.  We may also
consider the affine spaces $S_x^j$ as sets of states over a larger finite
subset of $\mathbb{Z}^d$ containing $x$.

Now, given a finite subset $A\subseteq\mathbb{Z}^d$, with
$A=\{x_1,\ldots,x_n\}$, the set of lattice states which maps to a particular
state $\ket{j}=\ket{j_1,\ldots,j_n}$ on the lattice subset $A$ will be
$$S_A^j=\bigcap_{k=1}^nS_{x_k}^{j_k},$$
where the sets $S_{x_k}^{j_k}$ are considered as sets of states over the
lattice subset $\mathcal{N}(A)$.

By selecting a basis for the set of states over the lattice points in
$\mathcal{N}(A)\setminus A$ for each $j$, it is possible to find a unitary
operator $U_{\mathcal{N}(A)}$ that makes the
appropriate map, so any vector in $S_A^j$ is properly mapped to $\ket{j}$.  In
addition, note that in order for $U_A$ to satisfy the consistency condition,
vectors outside $S_A^j$ must not be mapped
to $\ket{j}$.  Since every vector in $S_A^j$ must be mapped to something, it
must be $\ket{j}$.  Thus every operator $U$ that satisfies the consistency
condition can be constructed in this way.  Also, note that while we have many
choices for any particular operator $U_A$, every choice yields the same result
after the partial trace.  Next, we will show these operators are always
consistent.

In order to be a proper global transition operator, $U$ must preserve the
consistency condition.  That is, given a state $\rho$, and given that
$\sigma=U\rho$, we must have
$$\sigma_A=\tr_{B\setminus A}(\sigma_B)$$
for every finite subset $B\subseteq\mathbb{Z}^d$ and for any $A\subseteq B$.
In other words, we must have
\begin{multline*}
\tr_{\mathcal{N}(A)\setminus A}\left(U_{\mathcal{N}(A)}\rho_{\mathcal{N}(A)}
U^{\dagger}_{\mathcal{N}(A)}\right)
=\\
\tr_{B\setminus A}\tr_{\mathcal{N}(B)\setminus B}\left(U_{\mathcal{N}(B)}
\rho_{\mathcal{N}(B)}U^{\dagger}_{\mathcal{N}(B)}\right),
\end{multline*}
given that $\tr_{\mathcal{N}(B)\setminus\mathcal{N}(A)}\rho_{\mathcal{N}(B)}
=\rho_{\mathcal{N}(A)}$

Since both $U_{\mathcal{N}(A)}$, $U_{\mathcal{N}(B)}$ and the partial trace
are linear functions, it suffices to show that this relation holds for
operators of the form $\rho_{\mathcal{N}(B)}=\ket{\psi}\!\bra{\phi}$,
where $\ket{\psi}\in S_B^j$ and $\ket{\phi}\in S_B^k$ for computational basis
states $\ket{j}$ and $\ket{k}$ over $B$.  But by definition, we also have
$\ket{\psi}\in S_A^j$ and $\ket{\phi}\in S_A^k$, and so both sides of the
above identity equal
$\tr_{B\setminus A}\ket{j}\!\bra{k}.$

Therefore, an operator $U$ is valid if and only if it is the same as the one
constructed by the local operator, up to unitary operators acting on
$\mathcal{N}(A)\setminus A$.

\section{QCA Models}

Now that we have shown our general QCA model, we can introduce particular
varieties of QCA types as instances of our model. The first one we introduce
is the Margolus QCA, or MQCA.

\subsection{Margolus QCA}

\begin{figure}
  \begin{center}
  \includegraphics[scale=0.4]{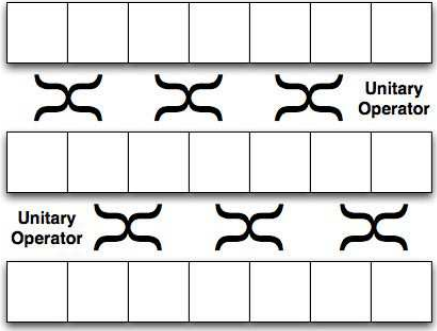}
  \end{center}
\caption{\label{mqca} \emph{Margolus QCA:} Similar to the classical reversible
CA, the lattice is partitioned using two different tiling sets. Each tile is
then acted on with a unitary operator.}
\end{figure}

The MQCA is similar to its classical counterpart, in which we define a tiling
of the lattice, and apply the same unitary operation to each tile. The
tiling is then changed in such a way that each tile of one tiling overlaps
with at least two tiles of the original one.  The two tilings need not have
equally shaped tiles, but must have the same period with respect to the
lattice (see Figure \ref{mqca}).

 Formally, we state:
 
\begin{definition}
A Margolus partitioning is a pair of array partitions $k$ and $r$, dividing
the lattice into collections of identical, finite, disjoint, and uniformly
arranged \emph{blocks}, such that each block of one partition overlaps with at
least two from the other. A Margolus Quantum Cellular Automata (MQCA) is
defined by lattice $S$ a Margolus partitioning of $S$ and a pair $U, V$ of
unitary transformations, each one acting on the blocks of one of the
partitionings.
\end{definition}

It is easy to see that the MQCA follows the consistency conditions outlined
above. An interesting example of an MQCA is the multi-particle quantum walk.

\begin{example}
Let $S$ be $\C^\Z$, the integer line. The partitionings $p$ and $s$ both divide
the lattice into blocks of two qubits, each block overlapping one qubit from
one block in the other partitioning, as in Figure \ref{mqca}.  Let

\[ u = v =  \left(
\begin{array}{cccc}
1 & 0 & 0 & 0\\
0 & \frac{1 + i}{2} &  \frac{-1 + i}{2} & 0  \\
0 & \frac{-1 + i}{2} &  \frac{1 + i}{2} & 0 \\ 
0 & 0 & 0 & 1\\
\end{array}
\right).
\]
\end{example}

For this example, in the computational basis, we understand a $1$ to represent
the presence of a particle, and zero to represent the abscence. The overall
effect is that any particles in the lattice slowly get diffused over time.
There is no interaction among particles.

	\begin{figure}
  \begin{center}
  \includegraphics[scale=0.36]{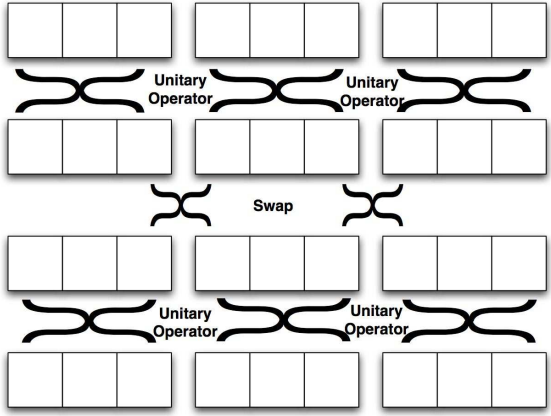}
  \end{center}
\caption{\label{pqca} \emph{Partitioned QCA:} In a PQCA each cell is partitioned into 3 \emph{`subcells'}: \emph{left, centre, and right}.. A unitary operator is applied to each cell at odd time steps, and at even time steps the cells exchange subcells, it's right subcell is exchanged with the left subcell of the right neighbour, and the left subcell is exchanged with the right subcell of the left neighbour.}
\end{figure}

Both van Dam \cite{vandam} and Watrous \cite{watrous} consider
\emph{partitioned} QCA or PQCA. A PQCA is shown in detail if figure \ref{pqca}. This scheme is simply a particular kind of
Margolus partitioning; one of the Margolus partitioning operators is simply a swap.

It should be pointed out, though, that the van Dam and
Watrous model differs greatly from the one presented here. In their model, the
quantum state of QCA is in a superposition of \emph{classical} configurations
of the lattice. This leads to a very different definition of the evolution
operator as well, which runs into problems as outlined in
\cite{schumacher}. Type-II quantum \cite{love1, love2}, computers can be regarded as particular instances of the PQCA scheme, except that the exchange procedure is done non-coherently.

The example we chose also has a historical significance. Quantum walks on a
lattice have been extensively studied before (see for example \cite{walks} for an overview)), and in particular a QCA
quantum walk has been analyzed in \cite{boghosian}. It is of note that one can take
the continuous limit of the walk we outlined above, i.e. the limit when the
\emph{space} between lattice cells and the duration of the time step both go
to zero. Doing so gives the Schr\"{o}dinger equation for a particle in free
motion \cite{boghosian}. This works for lattices of any dimension. By taking
the limit in a slightly different  way one can obtain the Dirac equation;
although this works only for one dimensional systems at the moment
\cite{meyer1}.

Also, in \cite{schumacher} Schumacher and Werner show that MQCA are \emph{universal} QCA, in the sense that all QCA (as defined in their general scheme) can be reduced to MQCA.

More recently, Raussendorf shows how to build a universal quantum computer
using a Margolus style quantum cellular automata \cite{raussendorf1}. For
these reasons, implementing a MQCA on a physical device would be of great
usefulness. We explore this in our next section.

\subsection{Colored QCA}

In 1993 S. Lloyd introduced an example of a (time-inhomogenous) CQCAs, which
we will call Spin-chain QCA. He introduced this model as a proposal for a
feasible quantum computer.  Consider a one-dimensional chain of spin
$1/2$ systems, such as a polymer, with three different \emph{species}, i.e
$ABC ABC\ldots$

Suppose that the nearest-neighbour interactions are given by (arbitrary)
Hamitlonians $H_{AB}$, $H_{BC}$, $H_{CA}$.  In the computational basis, the
diagonal terms of this Hamiltonian shift the energy levels of each cell as a
function of the energy levels of its neighbours \cite{lloyd93}.

The resonant frequency $\omega_A$ can then take the value $\omega^A_{00}$,
$\omega^A_{01}$, $\omega^A_{10}$, or $\omega^A_{11}$, depending on whether its
$C$ and $B$ neighbours are in the states $0$ and $0$, $0$ and $1$, $1$ and
$0$, or  $1$ and $1$.

Hence, chain link $A$ with neighbours $C$ and $B$ has resonant frequency
$\omega_A$ with distinguishable diagonal terms $\omega^A_{00}$,
$\omega^A_{01}$, $\omega^A_{10}$,  and $\omega^A_{11}$. If these are all
different, then transitions on species $A$ spins can be done selectively
depending on the value $k$. For instance, by applying a $\pi$ pulse with
frequency $\omega^A_{00}$ all species $A$ lattice points whose both neighbours
are in the state $0$ will be flipped.

It is also possible to apply, using the same techniques, any arbitrary
single-qubit gate to all spins of the same species and to apply any two qubit
gate on all pairs $A,B$ or $C,A$ or $B,C$.

We introduce a new general QCA model, which we call \emph{Coloured} QCA
(CQCA). This model is a generalization and abstraction of Lloyd's scheme.

  \begin{figure}
 \begin{center}
  \includegraphics[scale=0.3]{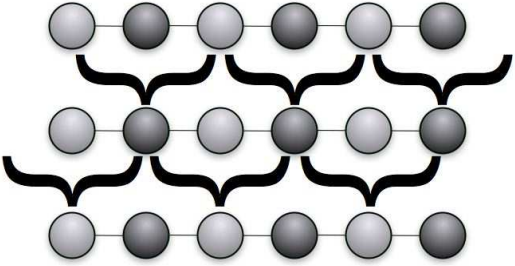}
    \end{center}

\caption{\label{cqca} \emph{Colored QCA} This is a representation of a
one-dimensional CQCA, with time going downward. The two shades of gray
differentiate between the two species in this CQCA. The brackets represent the
controlled unitaries: at each step either the light or dark grey sites are
acted on with unitary operators that depend on their left and right neighbour
values.}
\end{figure}

In a CQCA, each lattice point is assigned a colour in a checkerboard fashion.
At each time step only points of a certain colour are updated with a unitary
dependant on their neighbour's values. Neighbours of the same colour are {not}
distinguishable.  
  
A graphical representation is shown in Figure \ref{cqca}, and a formal
definition is given below.
   
\begin{definition}
A correct $k$-colouring for a lattice is a periodic mapping $C$ from lattice
points to $\{0, \ldots, k-1 \}$, such that no two neighbours have the same
colour. Fix a single qubit  observable $\sigma$. For each lattice cell $t$ its
$l$-field is the value $\sum_{r \in N(t), C(r) = l} \bra{\psi} \sigma^{(r)}
\ket{\psi}$.  A field-controlled unitary $u$ is such that it is only applied
to a lattice point if its l-field, for $l \in  \{0, \ldots, k-1 \}$ has a
particular set of values.

A Coloured QCA or CQCA is defined by a neighbouring $N$, colouring $C$, and a
field-controlled unitary $u_l$ for each colour $l \in \{0, \ldots, k-1 \}$.
\end{definition}

Again, it is not hard to see that CQCA adhere to the consistency constraints
outlined in our generic model.

Several proposal for implementation of quantum computers can be seen as
instances of the CQCA model. Obviously, Lloyd's pulse-drvien quantum computers
are one example. Benjamin's proposal \cite{benjamin} is an example of 2-CQCA
scheme that achieves universality.  K. G. H. Vollbrecht and J. I. Cirac
\cite{cirac} present a 1-CQCA (each lattice site being a qudit with $d = 4$),
that also achieves universality.

\begin{example} An example of a proper CQCA is the 1-D quantum walk
CQCA. As in Example 1 above, let $S$ be $\C^\Z$, the integer line. The
colouring scheme uses four colours $\{c_i\}_{0 \leq i \leq 3}$ such that
position $x_i$ on the lattice has colour $i \mod 4$. 

The evolution operator is as follows. First, $c_1$ is given a $\pi$
z-rotation conditioned on $c_0$ being in $\ket{1}$. Then $c_0$ is given a
$\pi/2$ z-rotation conditioned on $c_1$ being in $\ket{1}$. Then we repeat the
$\pi$ z-rotation on $c_1$ conditioned on $c_0$ being in $\ket{1}$. 
We can see that this procedure is equivalent to doing a
square-root-of-swap on all pairs $c_0$-$c_1$. We do the same procedure on
colours $c_2$-$c_3$, and then on both pairs $c_1$-$c_2$ and $c_3$-$c_0$.

The above construction gives the exact dynamics as the MQCA of Example 1
above, namely, a quantum walk on the line.

\end{example}

From the example above, one might deduce that any dynamic given in the MQCA
model can also be constructed using CQCA. This is in fact true, and is proven
below.

\begin{theorem}
For every MQCA there exists a CQCA, and for every CQCA there exists and MQCA
that has the exact same dynamics.
\end{theorem}

\begin{proof}
For this proof we rely on the fact that arbitrary single qubit operations,
coupled with c-NOT gates on adjacent qubits (for a connected topology) is
universal for quantum computation.

We proceed constructively. Given an MQCA $M$, with partitions $s$, $t$, we can
build a CQCA $C$ that has the exact same dynamics as follows.

The lattice of $M$ will be the same lattice for $S$. Without loss of
generality suppose that the partition $s$ has more lattice sites per block
than $t$. Let $k$ be the number of sites in each block of $s$. We construct
our $CQCA$ with $2 s$ colours. The colour mapping is such that each partition
block of $M$ is represented by $s$ distinct colours in $C$ (each block having
the exact same colours as each other block in the partition), and two
neighbouring block has two similar colours.

Now, we can simulate any arbitrary unitary acting on a block of $M$ using only
colour-controlled operations in $C$: since each spin in the block has a
distinct colour, they are all individually addressable. Moreover, we can apply
c-NOT gates between any two neighbouring spins. Operations on $C$ will be
repeated periodically over the lattice, due to the repeating colour scheme,
and hence it is important that block boundaries coincide with colour
periodicity.

Also we use $2 s$ colours instead $s$, so it is necessary to repeat the
operations on $C$ for alternating blocks of $M$. This is necessary to ensure
that we can isolate each block in order to perform arbitrary operations on it.

\end{proof}

The importance of the above theorem, beyond mathematical curiousity, is that
it allows us a simple way to implement MQCA algorithms on physical devices.
 
\subsection{Continuous-Time QCA}

Another model we present is the \emph{Continuous Time} QCA (CTQCA).
  
\begin{figure}
\begin{center}
\includegraphics[scale=0.3]{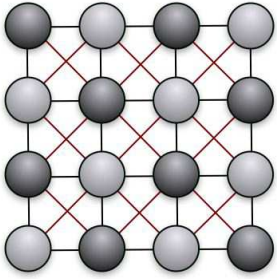}
\end{center}
\caption{\label{ctqca} \emph{Continuous Time QCA:} We have a crystal with two
types of nuclei $A$ and $B$, one represented as light gray spheres, the other
dark gray.  The lines connecting the spheres represent the nearest neighbour
couplings, with different colour lines representing different coupling
Hamiltonians. Notice that the coupling Hamiltonian depends only on the colour
of the coupled spins.}
\end{figure}

A CTQCA is similar to a CQCA in that all lattice points are coloured. However,
instead of unitary operators applied in discrete timesteps, the system
evolves continuously according to a Hamiltonian described only by nearest
neighbour couplings.

\begin{definition}
Let $C$ be any colouring, that is any periodic mapping from lattice points to
$\{0, \ldots, k-1 \}$, but not necessarily `correct' in the sense of
Definition 2 above.

Every pair of neighbouring lattice points $i,j$ has a coupling Hamiltonian
$H_{i,j}(t)$ that depends only on the colour of the two points. For a given
region of the lattice, the Hamiltonian of that region is simply
\[ H(t) \sum_{i<j}  H_{i,j}(t).\]

The evolution of the CTQCA over a given time period $t$ is given by
\[ U[t] = \int e^{i H(t) } dt.\]
\end{definition}

A good example of a CTQCA is the diffusion automata as defined below.

\begin{example} Again, as in example 1 above 
Let $S$ be $\C^\Z$, the integer line. The colouring scheme requires only one
colour, and each spin has only two neighbours, one directly to its left, and to
its right. The coupling Hamiltonian is
\[ H = \sum_i \sigma_+^{(i)} \sigma_+^{(i+1)} +  \sigma_-^{(i)} \sigma_-^{(i+1)}\]

In solid-state NMR this is called the flip-flop coupling, since it has the
effect of flipping two contiguous spins from $\ket{01}$ to $\ket{10}$. 

\end{example}

The evolution of this CTQCA is similar to the quantum walks of Examples 1 and
2 above.  Again this is not coincidental.

An important result is that CTQCA is, again, an equivalent computational model
to the other QCA models presented above.
 
\begin{theorem}
Given, a CQCA there is a CTQCA that has the same dynamics.
\end{theorem}
 
\begin{proof}
This construction is very simple. Given a a CQCA $C$ we can construct a CTQCA
$T$ with the same lattice and same colour-scheme. We give a time-dependant
Hamiltonian $H$ that changes at discrete time-steps of duration $t$ so that $U
= e^{i t H}$ where $U$ is the unitary of the CQCA at the given timestep.
 
\end{proof}
 
The converse is slightly more complicated, and can only be done in an approximate sense,

\begin{theorem}
Let $\Delta T$ be the time it takes to do one time step. Then, a CQCA $C$ can
approximate a CTQCA $T$ when $\Delta T \rightarrow 0$. 
\end{theorem}

\begin{proof}
Continuous time QCA, as opposed to CQCA do not have the restriction that
neighbours cannot be of the same colour. Assuming that the colouring is
correct for the given CTQCA $T$, then we proceed as follows.
 
Let $C$ have the same colour scheme as $T$. 

For any pair of colours $t_i$ and $t_j$ let $H_{i,j}$ be the coupling
Hamiltonian for neighbours of that colour. Let $U_{i,j}[t] = e^{- i t H}$.
Now, $U_{i,j}[t] $ acts only on two spins, and hence can be described as a
series of  controlled operations, $U_{i,j}^{(i,1)}[t]$, $U_{i,j}^{(j,1)}[t]$,
$U_{i,j}^{(i,2)}[t]$, $U_{i,j}^{(i,1)}[t]\ldots$ where $U_{i,j}^{(i,s)}[t] $
is an operation on the spin of colour $i$ controlled by the spin $j$, and so
forth. These $U_{i,j}^{(r,s)}[t]$ become the controlled operators in our CQCA
$C$. By letting $t \leftarrow 0$ the evolution of $C$ approximates the
evolution of the CTQCA $T$.
 
If $T$ has neighbours of the same colour, we simply break any colour that has
neighbouring spins into two or more colours, such that there are no longer
neighbouring spins of the same colour. All the colours created this way will
have the same coupling Hamiltonian to each other as the original colour had
with itself. We can then apply the procedure outlined above.
 
\end{proof}

\section{Conclusions}

In closing we wish to give a summary of the contributions of this paper. First
we gave a simple scheme for QCA that is general enough to encompass previous
models of QCA, as well as other phenomenae studied under different names.

We gave several specific models, under this scheme, and we showed the
relationship among this models. We posit that our model is not overly general,
that is it describes all well formed QCA style phenomana and \emph{nothing
else}. This latter statement, though, is posited without proof.

\section{Acknowledgements}

Research for this paper was supported in part by ARDA, ORDCF, 
CFI, MITACS, and CIAR.

\nocite{*}
\bibliographystyle{h-physrev3}
\bibliography{qca1}

\end{document}